\documentclass[runningheads]{llncs}
\PassOptionsToPackage{table}{xcolor} 
\usepackage[T1]{fontenc}
\usepackage{graphicx}
\usepackage{longtable}
\usepackage{booktabs,tabularx}
\usepackage{array}
\usepackage{multirow}
\usepackage{comment}
\usepackage{amsmath,amssymb}

\usepackage{orcidlink}
\usepackage{hyperref}

\usepackage{subcaption}

\usepackage[per-mode=symbol]{siunitx}

\newcommand{\unc}{\tiny}
\usepackage{makecell}

\setcellgapes{1.5pt}

\usepackage{lipsum}

\def\benchmarkAcronym{CroQS}%
\def\taskName{cross-modal query suggestion} 
\def\taskNameCapit{Cross-modal query suggestion}

\def\specificityMetricCap{Cluster Specificity} 
\def\clipQueryCap{CLIP Query}

\definecolor{verylightgray}{gray}{0.95} 

\begin{document}
%
\title{
Maybe you are looking for CroQS \includegraphics[height=1em]{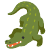} \\Cross-modal Query Suggestion for Text-to-Image Retrieval
}
\titlerunning{\includegraphics[height=1em]{crocodile-emoji.pdf} CroQS: Cross-modal Query Suggestion for Text-to-Image Retrieval}

\author{
Giacomo Pacini \inst{1,2}\orcidlink{0009-0007-7745-4456} \and
Fabio Carrara \inst{1} \orcidlink{0000-0001-5014-5089} \and
Nicola Messina \inst{1} \orcidlink{0000-0003-3011-2487} \and \\
Nicola Tonellotto \inst{1,2} \orcidlink{0000-0002-7427-1001} \and
Giuseppe Amato \inst{1} \orcidlink{0000-0003-0171-4315} \and
Fabrizio Falchi \inst{1} \orcidlink{0000-0001-6258-5313}
}
\authorrunning{G. Pacini et al.}
%
\institute{
 CNR-ISTI, Pisa, Italy \email{\{name.surname\}@isti.cnr.it} \and
 University of Pisa, Italy \email{\{name.surname\}@unipi.it}
}

\maketitle              
\begin{abstract}

Query suggestion, a technique widely adopted in information retrieval, enhances system interactivity and the browsing experience of document collections.
In cross-modal retrieval, many works have focused on retrieving relevant items from natural language queries, while few have explored query suggestion solutions. 
In this work, we address query suggestion in cross-modal retrieval, introducing a novel task that focuses on suggesting minimal textual modifications needed to explore visually consistent subsets of the collection, following the premise of ``Maybe you are looking for''.
To facilitate the evaluation and development of methods, we present a tailored benchmark named CroQS.
This dataset comprises initial queries, grouped result sets, and human-defined suggested queries for each group.
We establish dedicated metrics to rigorously evaluate the performance of various methods on this task, measuring representativeness, cluster specificity, and similarity of the suggested queries to the original ones.
Baseline methods from related fields, such as image captioning and content summarization, are adapted for this task to provide reference performance scores.
Although relatively far from human performance, our experiments reveal that both LLM-based and captioning-based methods achieve competitive results on CroQS, improving the recall on cluster specificity by more than 115\% and representativeness mAP by more than 52\% with respect to the initial query.
The dataset, the implementation of the baseline methods and the notebooks containing our experiments are available here: \href{https://paciosoft.com/CroQS-benchmark/}{paciosoft.com/CroQS-benchmark/}

\end{abstract}
\section{Introduction}

\begin{figure}[t]
    \centering
    \includegraphics[width=\textwidth]{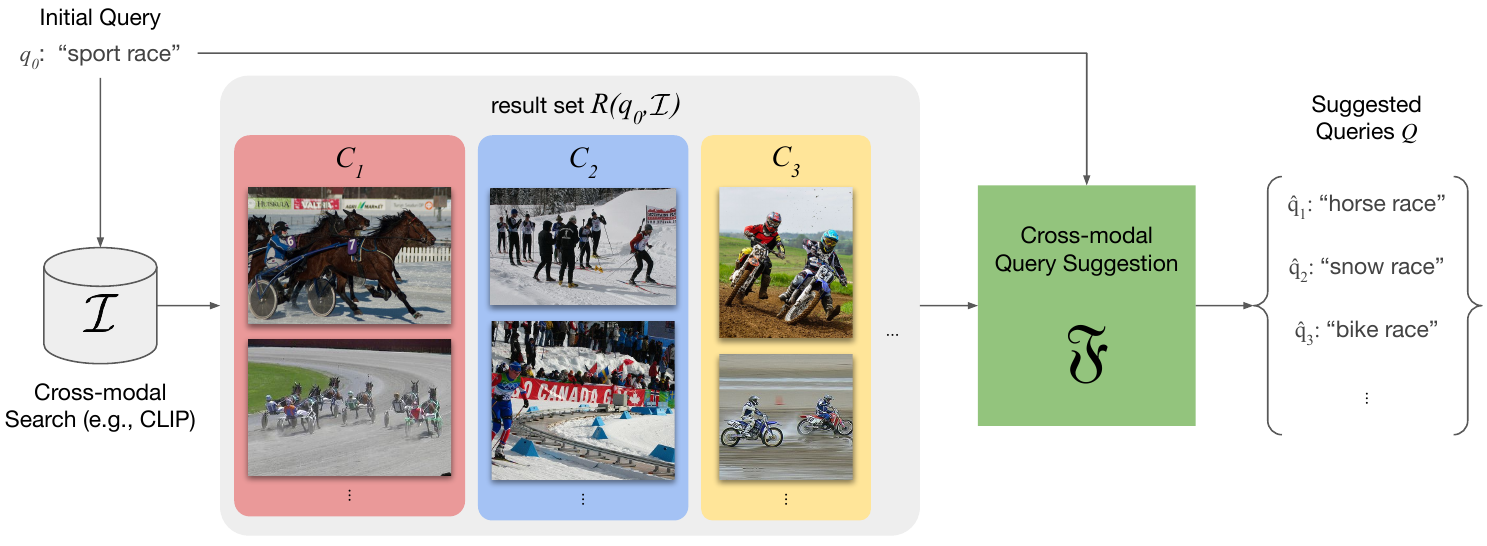}
    \caption{\textbf{Cross-modal Query Suggestion}. Given an initial query $q_0$ and an image collection $\mathcal{I}$, a cross-modal query suggestion system $\mathfrak{F}$ returns a set of query suggestions $\mathcal{Q}$ based on the visual content of the result set $R(q_0, \mathcal{I})$. Ideally, each suggestion $\hat{q}_i \in \mathcal{Q}$ should represent a semantically coherent group $C_i \subset R(q_0, \mathcal{I})$.}
    \label{fig:task-schema}
\end{figure}

Query formulation is a central challenge in Information Retrieval (IR), as it consists of aligning user queries with their actual information needs. 
Query expansion and query suggestion are two families of query reformulation approaches that aim to better align the search results with the user's information need \cite{ooi2015surveyQueryExpansionQuerySuggestion}.
The first one is generally automatic and modifies the IR system's internal representation of the original query.
Query suggestion, instead, is an interactive approach that proposes different queries to the user to refine the search.
Those techniques either base their reformulations on the initial search results --- as happens in Rocchio's algorithm \cite{rocchio1971relevance} with its Pseudo Relevance Feedback (PRF) --- or on 
external knowledge sources, such as thesauri, ontologies, or large-scale language models.
In the case of PRF, the system uses feedback from the initial set of retrieved documents to automatically adjust the query, while knowledge-based approaches leverage pre-defined relationships between terms to suggest more effective formulations.

Historically, IR and query reformulation approaches have been designed for document retrieval, where a textual query is employed to retrieve textual documents.
Therefore, methods, architectures, and datasets that define and solve this task mainly consider the text modality alone. 
Recently, \textit{cross-modal retrieval} and, in particular, \textit{text-to-image retrieval} is becoming the key to searching large image and video datasets, thanks to the recent development of powerful models that embed images and texts in the same semantic feature space \cite{messina2022aladin,messina2021fine,CLIPradford2021learning}.
In particular, CLIP \cite{CLIPradford2021learning} has simplified research in this area and enabled further possibilities. It embeds text and visual data into the same embedding space and can easily retrieve the most relevant images or videos by computing a maximum inner-product search with the textual query embedding. This model is nowadays the de facto standard in multimedia retrieval, and it has been largely employed in interactive video browsing competitions \cite{lokovc2023interactive,amato2024visione,gurrin2024introduction}.

Given the growing interest in cross-modal search applications, 
there is a pressing need to develop a tool that interactively suggests queries in cross-modal scenarios --- an area that is under-explored in the literature.
For this reason, in this work, we aim to explore the query suggestion task in the scenario where queries are provided in natural language, and the search database comprises a potentially large set of unannotated images.
We refer to this task as \emph{cross-modal query suggestion}, which consists of proposing suitable modifications to the initial user query based on visual clues present in the retrieved images.
Ideally, each suggestion should represent different aspects of the original result set and allow the user to interactively understand the semantic groups of images that populate the search collection.
Figure~\ref{fig:task-schema} shows a schematic overview of the task with an example.

The proposed task is challenging for many reasons.
In a cross-modal retrieval scenario, document collections are usually made of plain images that do not carry captions, textual descriptions, or any additional metadata of the documents.
Furthermore, 
the information to extract from an image document can change 
depending on the context.
For this reason, textual features statically provided with the images --- such as the captions of image-caption datasets --- are not effective for the query suggestion task since they lose some of the possible \textit{meanings} of an image.
To solve this problem, it is necessary to define a mechanism that, given an initial query and a semantic representation of the result images, constructs a suggested query.
Considering the novelty of this problem in cross-modal retrieval, it is also necessary to define a method and a benchmark for method comparison. 
To this aim, 
we introduce the \benchmarkAcronym{} benchmark.
We developed \benchmarkAcronym{} through a semi-automatic, human-supervised process.
We generated every semantic cluster of images from an initial query and a clustering step, further tuning the obtained sets through careful human judgment.
A human annotator also defined a reference query suggestion per group to have a solid ground truth.
In order to objectively evaluate different methods with the benchmark, we defined a set of metrics, and we adapted several baseline methods from related fields --- such as image captioning and content summarization --- to solve the task.

To summarize, the contributions of this paper are threefold:
\begin{itemize}
    \item We define and tackle the novel task of cross-modal query suggestion, which extends classical text-based query suggestion by directly incorporating latent high-level information extracted from the retrieved images.
    \item We provide the \benchmarkAcronym{} benchmark, accompanied by proper evaluation metrics, to quantitatively evaluate models on this new challenging task.
    \item We propose some reasonable baselines for this benchmark by adapting some captioning-based and LLM-based methods.
\end{itemize}

Our results show that the methods derived from captioning are more effective in building queries that are specific to their group of images, while LLM-based approaches are balanced throughout the properties and tend to build suggestions very similar to the initial queries.
\section{Related Works}
\label{related-works}

Cross-modal query suggestion is a task spanning different research domains.
Besides classical IR query suggestion, this task borrows from methods extracting image content as textual descriptions, thus relating to image group captioning and image captioning in general.
While image group captioning is about building a natural language sentence that describes a set of images, cross-modal query suggestion is about generating a more detailed query that better explains a group of query results.

\subsection{Query Suggestion}

In dense retrieval~\cite{khattab2020colbert}, 
query reformulation has a lower impact in improving retrieval metrics, while it can still be exploited to improve retrieval systems' interactivity and explorability.
Query suggestion provides to the user several queries that the system deems related to the user’s interest according to assumptions made by the retrieval system~\cite{ooi2015surveyQueryExpansionQuerySuggestion}.
Various works proposed solutions to build query suggestions based on large-scale graphs of queries and their click-through rates~\cite{cao2008contextAwareQuerySuggestionClickThrough,mei2008querySuggestionHittingTime}.
More recent approaches for solving this task in the textual domain employed RNN \cite{wu2018query} and transformer networks \cite{mustar2021study} or leveraged different learning schemas such as reinforcement learning \cite{bodigutla2021high}. Nowadays, the interest is shifting towards the use of Large Language Models (LLMs) \cite{baek2024knowledge,bacciu2024generating}, which show remarkable abilities in suggesting queries given the user intention or the search history as a prompt.
The work mostly resembling our objective is the one in Zha et al.~\cite{zha2009visualQuerySuggestion}, which proposed Visual Query Suggestion, a system for collections of captioned images that groups the result set of an initial query and suggests a reformulated query for each group.
However, they built the queries from the texts associated with the images leveraging a keyword selection algorithm, 
whereas our scenario operates under the assumption that no textual metadata is available for the images.

\subsection{Image Group Captioning}

The choice of image captioning as a foundational field is motivated by several factors.
In cross-modal query suggestion, the goal is to generate a suggested query that effectively represents a set of images while maintaining alignment with the initial query.
Similarly, image captioning methods aim to generate a descriptive sentence for a given image.
Although these two tasks are not identical, they share important commonalities in their input-output structures.
Both tasks start from visual inputs --- images --- and generate textual outputs.
This overlap suggests that techniques developed for image captioning may be adapted to handle cross-modal query suggestion.
The advent of CLIP~\cite{CLIPradford2021learning} 
introduced the possibility of representing texts and images in a cross-modal space and paved the way for various works in the image captioning field exploiting it.
Mokady et al. introduced ClipCap~\cite{mokady2021clipcap}, an architecture that, given a CLIP image embedding, generates a caption for it through the use of a mapping network and a large language model.
In 2023, Li et al.~\cite{li2023decap} introduced DeCap, which has the same goal but leverages a simpler architecture and achieves higher captioning results.

Wang et al. proposed a method to build image captions optimizing the \textit{distinctive} aspects of the input image with respect to the characteristics of a group of similar images \cite{wang2020compareDistinctiveCaptioningUsingSimilarImageSets,wang2021groupDistinctiveCaptioning,wang2022distinctiveCiderBtw}, but they still focus on captioning a single image.
The method proposed by Phueaksri et al.~\cite{phueaksri2024imageCollectionSummarizationSceneGraph} aims at summarizing a collection of images through their scene-graphs and an external knowledge graph into a scene graph of the whole collection.

N. Trieu et al.~\cite{trieu2020multiImageSummarization} proposed a transformer model for \textit{multi-image summarization} captioning a set of similar images in the specific domain of e-commerce product images.
Also Li et al.~\cite{li2020contextAwareGroupCaptioning} studied image group captioning and proposed a method that contrastively builds a caption for a target group of images comparing against an additional \textit{context group} of images, which serves as a reference point to distinguish the target group's unique characteristics. 
However, those methods do not consider an initial text (or query) to condition the generation of the caption as would be needed in cross-modal query suggestion.
\section{\taskNameCapit}

\subsection{Problem definition}

Let $q_0 \in \mathcal{T}$ be the initial query prompted by the user in natural language (being $\mathcal{T}$ the set of text strings) and $R(q_0, \mathcal{I}) \subset \mathcal{I}$ be the result set obtained by searching $\mathcal{I}$ with $q_0$, where $\mathcal{I} = \{I_i\}_{i=1}^D$ is the image collection.
Indicating with $2^\mathcal{X}$ the power set of $\mathcal{X}$ (the set of subsets of $\mathcal{X}$), 
we define a cross-modal query suggestion system $\mathfrak{F}: \mathcal{T} \times 2^\mathcal{I} \to 2^\mathcal{T}$ as
\begin{equation}
    \mathcal{Q} = \mathfrak{F}(q_0, R(q_0, \mathcal{I}))\,,
\end{equation}
where $\mathcal{Q} = \{\hat{q}_i \}_{i=1}^N \subset \mathcal{T}$ is the set of suggested queries, the output of the system.
All the suggested queries $\hat{q}_i$ 
should be \textit{variations of the same initial query $q_0$}, oriented to disambiguate and better explain the initial result set $R(q_0, \mathcal{I})$.

\subsection{Proposed Framework}

The core of the \taskName{} task revolves around formulating suggested queries based on the original query and the main concepts represented in the original result set.
We assume that an initial result set $R(q_0, \mathcal{I})$ contains one or more disjoint groups $\mathcal{C} = \{C_i\}_{i=1}^M \subseteq R(q_0, \mathcal{I}), C_i \cap C_j = \emptyset\; \forall i \neq j$ where each $C_i$ comprises images sharing semantic content with a higher specificity (e.g., ``a bike race'' or ``a skiing race'') than the original query $q_0$ (e.g., ``a sport race'').

In the proposed framework, the process begins with the user submitting an original query $q_0$, then the result set $R(q_0, \mathcal{I})$ provided by the retrieval system is partitioned into $M$ distinct semantic groups $C_i$ by a clustering algorithm. 
At this point, the \taskName{} system is tasked to generate a suggested query $\hat{q_i}$ for each group $C_i$. The suggestion system runs $M$ times, receiving as input $q_0$ and one semantic group at a time.
Finally, a \textit{user interface} presents the set of suggested queries $\{\hat{q_i}\}_{i=1}^M$ to the user for further exploration.

Under this assumption, we simplify the formulation as:
\begin{equation}
    \hat{q}_i = \mathfrak{F}(q_0, C_i)\,,
\end{equation}
where the \taskName{} system $\mathfrak{F}: \mathcal{T} \times 2^\mathcal{I} \to \mathcal{T}$ now operates on a set of semantically coherent images $C_i$ and produces a single suggested query $\hat{q}_i$ for that set.
We delegate the partitioning of the initial result set $R(q_0, \mathcal{I})$ into different semantic groups $\{C_1, C_2, \dots, C_M\}$ to off-the-shelf clustering algorithms operating on the semantic image representations of the collection (i.e., CLIP visual features in our experiments, the same adopted for the initial cross-modal search) and focus our investigation on how to implement $\mathfrak{F}$.

The proposed methodology, therefore, operates only after the semantic clusters have been established, focusing exclusively on generating refined queries based on the original input and the external clustering system's output. 

\section{Evaluation}

\subsection{Dataset}

Defining an objective way to compare different \taskName{} methods is not trivial. 
One of the most critical variables for a fair comparison is the definition of the clusters $\mathcal{C}$ for the given result set of images $R(q_0, \mathcal{I})$.
Two systems could come up with completely different and perhaps equally valid clusters, leading to difficult comparability of the results.

\begin{figure}[t]
    \centering
    \includegraphics[width=\textwidth]{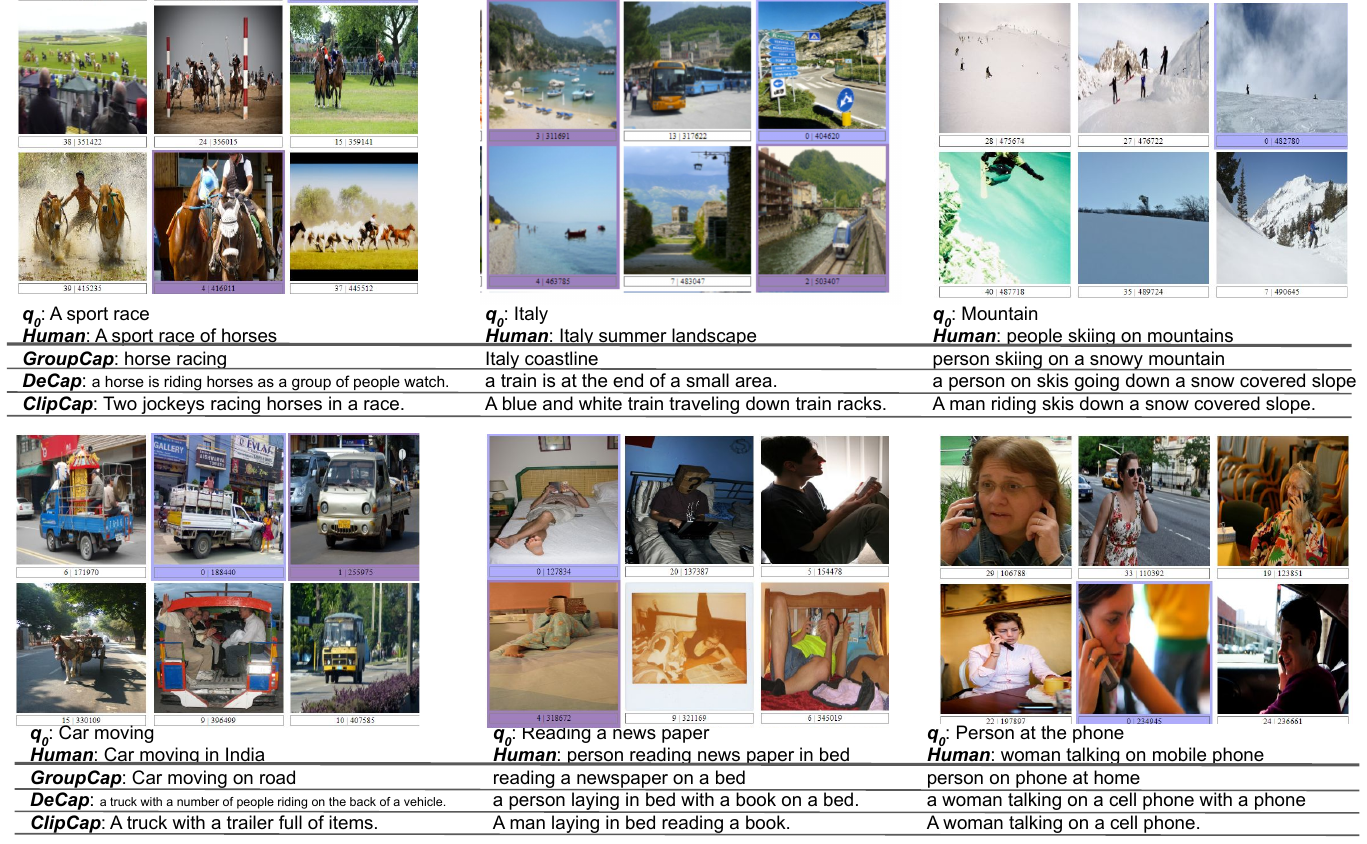}
    \includegraphics[width=\textwidth]{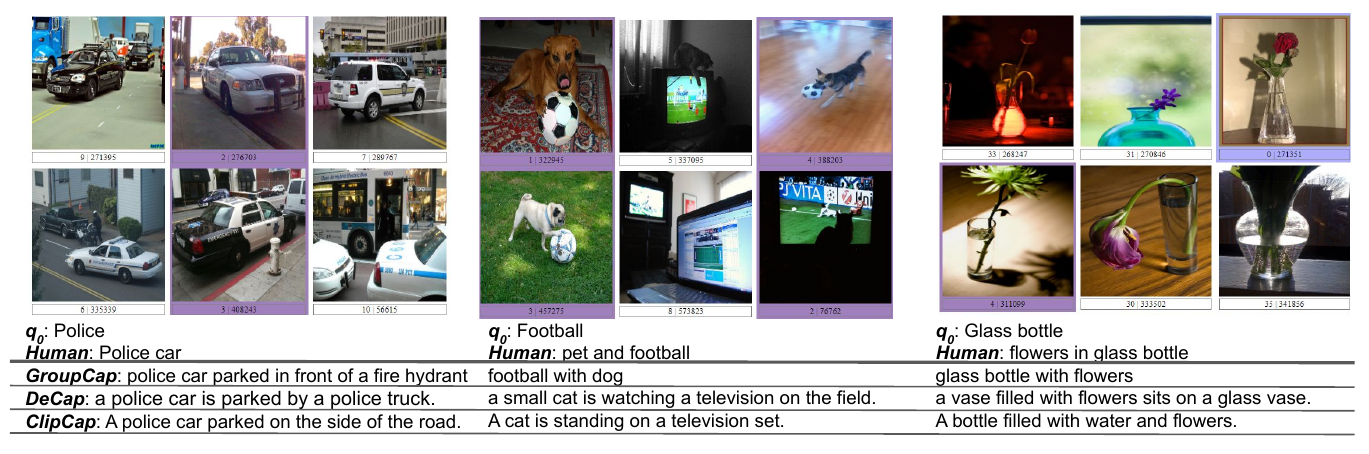}
    \caption{
    \textbf{CroQS samples and predictions}. Each panel reports a sample of a semantic cluster with its initial query $q_0$, its annotation, and the query suggested by the tested methods. %
    Representative images 
    (Eq. \ref{eq:representativeness}) 
    have a colored outline.%
    }
    \label{fig:qualitative-examples}
\end{figure}

Consequently, through manual work of collection exploration and query definition, we built a benchmark for \taskName{}, named \benchmarkAcronym{}, to factor out cluster variability in evaluating the generation of suggested queries.
\benchmarkAcronym{} comprises 50 initial textual queries. 
For each query, we manually validated the images in the result set and partitioned them into
a varying number of semantic clusters of images (min 2, max 10, 5.9 on average), totaling 295 clusters.
Each cluster is associated with a human-annotated suggestion for the target refined query.
We built \benchmarkAcronym{} on top of the \textit{train split} of COCO~\cite{lin2014microsoftCOCO}, a dataset of more than 118,000 captioned images.
The image captions provided by COCO only assisted us in defining the human-annotated suggestions and were otherwise discarded for our purposes.
Figure \ref{fig:qualitative-examples} shows samples from \benchmarkAcronym{}.

Employing \benchmarkAcronym{}, it is possible to compare different methods only by means of the suggested queries they generate. 
In the following, we present the properties to be evaluated and a set of metrics that we can exploit to measure them.

\subsection{Evaluation Metrics}

Due to the novelty of the \taskName{} task, defining the properties to be pursued is also necessary.
We distinguished three main desired properties of the generated suggested queries: 
\textbf{\specificityMetricCap{}}, \textbf{Representativeness}, and \textbf{Similarity to Original Query}.

\subsubsection{\specificityMetricCap{}} measures how the suggested query $\hat{q}_i$ describes the images belonging to its cluster $C_i$. In particular, it must highlight the properties that distinguish the images in $C_i$ from those in other clusters $C_j\,, i \neq j$.
To measure it, we quantify how well the suggested query $\hat{q}_i$ recalls the element of $C_i$ among the initial result set $R(q_0, \mathcal{I})$.
Specifically, we compute the following:
\begin{equation}
\text{Recall}_{\text{Cluster}} = \frac{|C_i \cap 
R_{[:k]}(\hat{q}_i, R(q_0, \mathcal{I}))%
|}{k}\,,
\end{equation}
where $C_i$ acts as the set of relevant documents, $R_{[:k]}(\hat{q}_i, R(q_0, \mathcal{I}))$ are the top-k elements of $R(q_0, \mathcal{I})$ when sorted by the suggested query $\hat{q}_i$ and act as the search collection, and
$k = |C_i|$.
A $\text{Recall}_{\text{Cluster}}$ of $1$ is obtained when all elements of $C_i$ are ranked before images belonging to other clusters.

\subsubsection{Representativeness} of the suggested query $\hat{q}_i$ for a group of results $C_i$ indicates how much $\hat{q}_i$ better represents the images in $C_i$ than the original query $q_0$.
To measure Representativeness, we look at how results in $C_i$ are ranked when searching with the suggested query $\hat{q}_i$ over the whole collection $\mathcal{I}$.
We compute standard retrieval metrics, i.e., Mean Average Precision (MAP), Normalized Discounted Cumulative Gain (NDCG), and Recall, 
considering the images belonging to the starting group of images $C_i$ as relevant.
\begin{equation}
\text{Recall} = \frac{|C_i \cap R(\hat{q}_i, \mathcal{I})|}{|C_i|}
\end{equation}
Then, while Representativeness Recall exclusively measures the ability of a method to retrieve the images of its cluster from the whole collection, NDCG and MAP also measure if those images are placed among the first results.

A suggested query can be specific but not representative, and vice versa.
Consider the case where the query $q_0$ is "\textit{A dog running}", the current cluster $C_i$ contains images of \textit{black} dogs running, and the result set $R(q_0, \mathcal{I})$ shows images of dogs running outside.
A suggested query $\hat{q}_i$ like "\textit{black}" could be specific but not representative. 
Conversely, if $\hat{q}_i$ is "\textit{dog running outside}", it would be representative but not specific at all.

\subsubsection{Similarity to the original query} measures how much the suggested query $\hat{q}_i$ diverges from the original query $q_0$, as the task requires suggesting corrections to a user-written query rather than re-writing it from scratch.
It is interesting to evaluate both the syntactic and the semantic similarity of the suggested queries.
For syntactic similarity, we adopted Jaccard score $\text{Jaccard}(q_0, \hat{q}_i)$, while for semantic similarity, we propose to adopt a similarity measure computed in an embedding space, such as the CLIP textual space $\text{CLIP}(q_0, \hat{q}_i)$. We refer to this latter metric as \clipQueryCap{}.

\section{Experiments}

\subsection{Baseline Methods}

We propose two baseline methods to tackle the \taskName{} task inspired by related fields --- image captioning and content summarization.

\begin{figure}[t]
     \centering
     \begin{subfigure}[t]{0.39\textwidth}
         \centering
         \includegraphics[width=\linewidth]{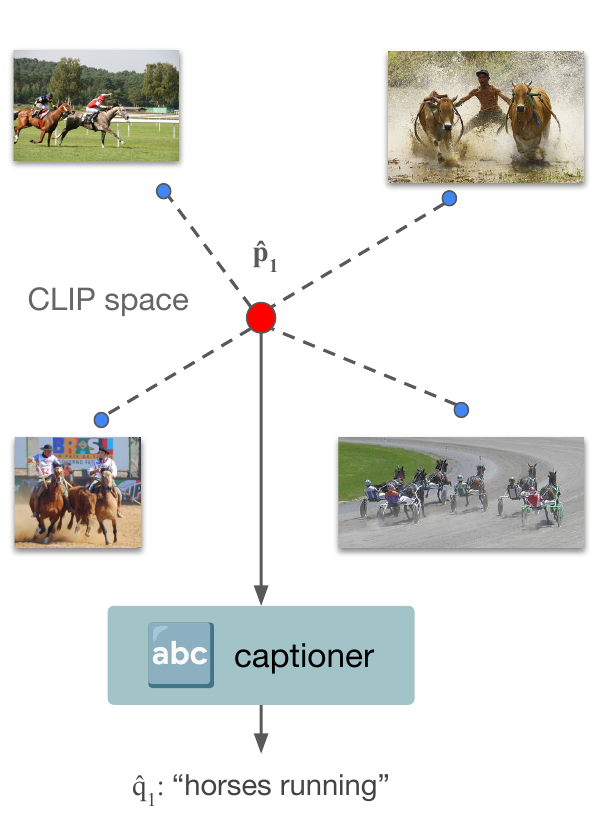}
         \caption{
         \textbf{Prototype Captioning}: 
         the suggested query is obtained by a captioning model (\textbf{ClipCap} or \textbf{DeCap}) applied to a cluster prototype in CLIP image space.}
         \label{fig:qe-from-captioning-clip-centroid}
     \end{subfigure}
     \hfill
     \begin{subfigure}[t]{0.54\textwidth}
         \centering
         \includegraphics[width=\linewidth]{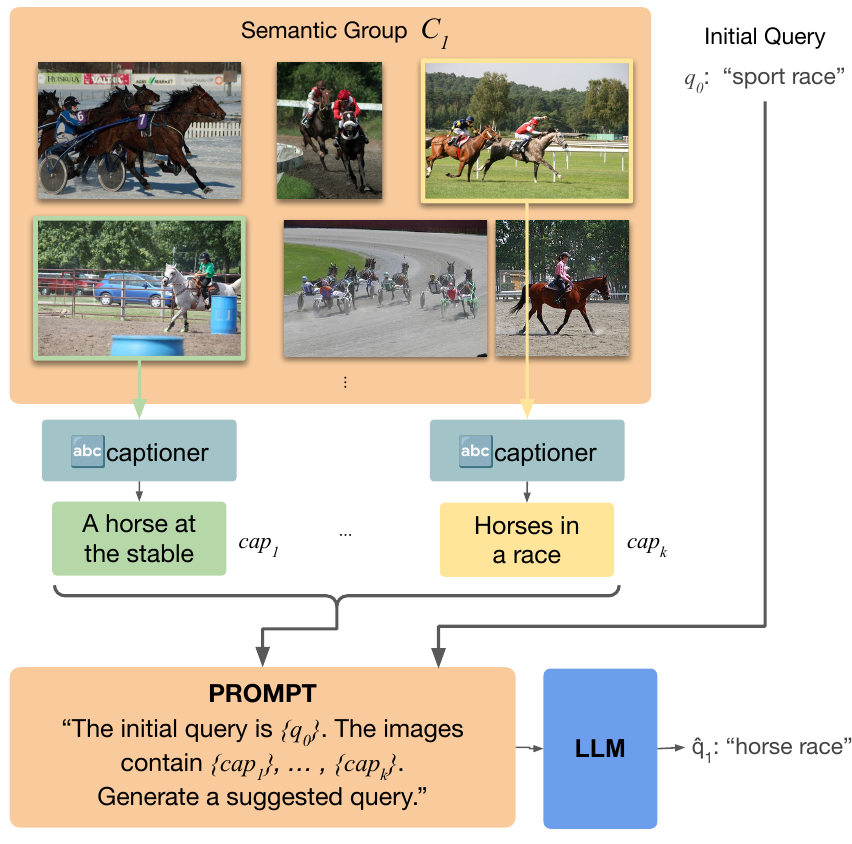}
         \caption{
         \textbf{Captions Summarization}:
         an LLM receives $q_0$ together with the captions of the $k$ most representative images and generates a query suggestion. We refer to this method as \textbf{GroupCap}.}
         \label{fig:groupcap-llm-schema}
     \end{subfigure}
     \caption{Architectures of the baseline methods proposed.}
\end{figure}

\subsubsection{Prototype Captioning}
As we analyzed in Section \ref{related-works}, image captioning can be considered a strongly related field to \taskName{}. 
However, there are significant differences that must be addressed. While traditional image captioning methods are designed to take a single image as input and generate a descriptive sentence, \taskName{} methods work with a set of images and aim to generate a single suggested query that encompasses all of them.
To bridge this gap, we propose utilizing continuous and semantically meaningful representations of the images rather than relying on discrete descriptions of individual images. By representing a set of images in such a semantic space, we can compute a meaningful prototype point $\hat{p}_i$ representing an entire group $C_i$. 
In particular, the CLIP's image space \cite{CLIPradford2021learning} is able to place semantically similar images close, making it an ideal choice for this task.

Based on this approach, we selected two captioning models --- \textbf{ClipCap} \cite{mokady2021clipcap} and \textbf{DeCap} \cite{li2023decap} --- which are built on CLIP's semantic space. These models can be fed with a prototype point from the CLIP space, without the need for an associated image, enabling them to generate text that is representative of the entire group of images. The core idea is to select a prototype point 
that is close enough to each image in the group, as in the example of Figure \ref{fig:qe-from-captioning-clip-centroid}. This prototype is then decoded to generate a representative text for the image group.

However, this strategy has a main limitation: off-the-shelf captioning models cannot be conditioned or guided by the initial query $q_0$ and provide an independent caption that usually deviates from $q_0$. This is reflected in lower scores in terms of similarity to the original query.
To overcome this limitation, we modified the ClipCap architecture to make it \textit{query-aware}.
As explained by Mokady et al. 
\cite{mokady2021clipcap}, ClipCap is made of two main components: a network that maps a CLIP image embedding to a set of GPT2 tokens, and a transformer decoder --- GPT2 \cite{gpt2radford2019language} in particular --- which decodes that set of tokens to a natural language image caption.
This two-step architecture allowed us not only to build a tailored input as explained above but also to provide the initial query $q_0$ to the second module of the architecture.
We will refer to this model as $\text{ClipCap}^{q_0}$.

\subsubsection{Captions Summarization}
Given the significant advancements in large language models (LLMs) and their remarkable ability to generate and understand text, we decided to explore how these models could be adapted for the \taskName{} problem. 
We employed a few-shot learning approach. This method allows the model to grasp the task requirements and generate the desired text without needing extensive retraining.

To achieve this, we propose the \textbf{GroupCap} architecture shown in Figure \ref{fig:groupcap-llm-schema}, which is designed to generate query suggestions through a series of steps. First, an algorithm selects the most relevant images from the cluster. These images are then processed by an image captioning method, which generates captions for each one. A carefully crafted prompt is created to instruct the LLM on how to use the captions and the initial query to generate a suggested query. The LLM then processes the formatted prompt and produces the final query suggestion.
This approach leverages the ability of modern LLMs to effectively understand and generalize tasks using few-shot prompts, making it an ideal method for adapting LLMs to the \taskName{} task.

\subsection{Quantitative Results}

This subsection reports the results obtained over our benchmark dataset by the adapted baseline methods.
For reference, the scores obtained by the human-annotated query suggestions are also reported.
Each score is computed as the Macro Average of the scores obtained on each cluster, for each initial query $q_0$.
Representativeness scores are calculated based on the top 100 documents of the result sets for the evaluated suggestions, and NDCG is assessed at rank 10.

\begin{table}
    \centering
    \caption{Comparison of the best \taskName{} configurations for each architecture. We report macro-averaged mean and and standard deviation.
    }
    \label{table:best-methods-scores-all-metrics}
    \resizebox{\textwidth}{!}{
    \begin{tabular}{
        l 
        c 
        c 
        c 
        c 
        c 
        c 
    }
    \toprule
    \textbf{} & \multicolumn{2}{c}{\textbf{Similarity to $q_0$}} & \multicolumn{1}{c}{\textbf{\specificityMetricCap}} & \multicolumn{3}{c}{\textbf{Representativeness}} \\
    \cmidrule(lr){2-3}\cmidrule(lr){4-4}\cmidrule(lr){5-7}
    \textbf{Method} & \makecell{\textbf{CLIP}} & \textbf{Jaccard} & \makecell{\textbf{Recall} \\ \textbf{Cluster}} & \makecell{\textbf{Recall}} & \textbf{NDCG} & \textbf{MAP} \\ 
    \midrule
    \rowcolor{verylightgray} $q_0$ & 1.00 \unc $\pm$ 0.00 & 1.00 \unc $\pm$ 0.00 & 0.19 \unc $\pm$ 0.07  & 0.52 \unc $\pm$ 0.03 & 0.18 \unc $\pm$ 0.05 & 0.21 \unc $\pm$ 0.05 \\
    \rowcolor{verylightgray} human & 0.87 \unc $\pm$ 0.03 & 0.44 \unc $\pm$ 0.11 & 0.59 \unc $\pm$ 0.13  & 0.62 \unc $\pm$ 0.12 & 0.52 \unc $\pm$ 0.14 & 0.45 \unc $\pm$ 0.11 \\
    \midrule
    ClipCap~\cite{mokady2021clipcap} & 0.74 \unc $\pm$ 0.06 & 0.17 \unc $\pm$ 0.11 & \textbf{0.54 \unc $\pm$ 0.16} & 0.40 \unc $\pm$ 0.15 & 0.32 \unc $\pm$ 0.15 & 0.28 \unc $\pm$ 0.12 \\
    DeCap~\cite{li2023decap} & 0.77 \unc $\pm$ 0.06 & 0.16 \unc $\pm$ 0.12 & 0.53 \unc $\pm$ 0.18 & 0.40 \unc $\pm$ 0.15 & 0.31 \unc $\pm$ 0.15 & 0.28 \unc $\pm$ 0.12 \\
    GroupCap & \textbf{0.90 \unc $\pm$ 0.04} & \textbf{0.56 \unc $\pm$ 0.12} & 0.39 \unc $\pm$ 0.17 & \textbf{0.47 \unc $\pm$ 0.12} & \textbf{0.35 \unc $\pm$ 0.18} & \textbf{0.32 \unc $\pm$ 0.14} \\
    \bottomrule
    \end{tabular}
    }
\end{table}

Table \ref{table:best-methods-scores-all-metrics} reports the metrics obtained by the best configuration of each architecture.
Human-annotated suggestions are getting the best results in all fields, except for similarity to the initial query, where the GroupCap model gets a higher average score.
The methods based on DeCap and ClipCap are biased towards the \specificityMetricCap{} metric, where those are getting the highest scores, while they rank last on the other metrics. This can be due to the original application field of these models, the captioning task, where the outputs need to be very specific and descriptive for the given image. 
This enabled them to effectively identify the characteristics that distinguish one cluster from the others for the same initial query, leading to high scores in \specificityMetricCap{}.
On the other hand, the higher level of detail of the suggestions generated by these methods often causes semantic shifts from the initial query, lowering their Representativeness scores.

The baseline methods achieve lower Representativeness Recall scores compared to $q_0$. 
By the way, it is important to note that the methods should find a trade-off between collection exploration and Representativeness.

\begin{table}
    \centering
    \caption{Comparison of different captioning-derived \taskName{} methods.
    Macro-averaged mean scores and associated standard deviations. 
    }
    \label{table:captioning-scores-all-metrics}
    \resizebox{\textwidth}{!}{
    \begin{tabular}{
        l 
        c 
        c 
        c 
        c 
        c 
        c 
    }
    \toprule
    \textbf{} & \multicolumn{2}{c}{\textbf{Similarity to $q_0$}} & \multicolumn{1}{c}{\textbf{\specificityMetricCap}} & \multicolumn{3}{c}{\textbf{Representativeness}} \\
    \cmidrule(lr){2-3}\cmidrule(lr){4-4}\cmidrule(lr){5-7}
    \textbf{Method} & \makecell{\textbf{CLIP}} & \textbf{Jaccard} & \makecell{\textbf{Recall} \\ \textbf{Cluster}} & \makecell{\textbf{Recall}} & \textbf{NDCG} & \textbf{MAP} \\ 
    \midrule
    DeCap$_\text{centroid}$ & \textbf{0.77 \unc $\pm$ 0.06} & \textbf{0.16 \unc $\pm$ 0.12} & \textbf{0.53 \unc $\pm$ 0.18}  & \textbf{0.40 \unc $\pm$ 0.15} & \textbf{0.31 \unc $\pm$ 0.15} & \textbf{0.28 \unc $\pm$ 0.12} \\
    DeCap$_\text{repr}$ & 0.73 \unc $\pm$ 0.06 & 0.13 \unc $\pm$ 0.10 & 0.52 \unc $\pm$ 0.16 & 0.36 \unc $\pm$ 0.15 & 0.30 \unc $\pm$ 0.16 & 0.27 \unc $\pm$ 0.12 \\
    \midrule
    ClipCap$_\text{centroid}$ & 0.72 \unc $\pm$ 0.06 & 0.14 \unc $\pm$ 0.10 & \textbf{0.55 \unc $\pm$ 0.17} & \textbf{0.40 \unc $\pm$ 0.15} & 0.30 \unc $\pm$ 0.15 & 0.27 \unc $\pm$ 0.11 \\
    ClipCap$_\text{repr}$ & 0.70 \unc $\pm$ 0.06 & 0.12 \unc $\pm$ 0.11 & 0.53 \unc $\pm$ 0.15 & 0.35 \unc $\pm$ 0.14 & 0.31 \unc $\pm$ 0.15 & 0.27 \unc $\pm$ 0.11 \\
    ClipCap$_\text{centroid}^{q_0}$ & \textbf{0.74 \unc $\pm$ 0.06} & \textbf{0.17 \unc $\pm$ 0.11} & 0.54 \unc $\pm$ 0.16 & \textbf{0.40 \unc $\pm$ 0.15} & \textbf{0.32 \unc $\pm$ 0.15} & \textbf{0.28 \unc $\pm$ 0.12} \\
    ClipCap$_\text{repr}^{q_0}$ & 0.72 \unc $\pm$ 0.07 & 0.15 \unc $\pm$ 0.10 & 0.52 \unc $\pm$ 0.16 & 0.37 \unc $\pm$ 0.16 & \textbf{0.32 \unc $\pm$ 0.15} & \textbf{0.28 \unc $\pm$ 0.12} \\
    
    \bottomrule
    \end{tabular}
    }
\end{table}

\begin{table}
    \centering
    \caption{Comparison of different GroupCap LLM-based \taskName{} methods. 
    We report macro-averaged mean and standard deviation.
    }
    \label{table:groupcap-scores-all-metrics}
    \resizebox{\textwidth}{!}{
    \begin{tabular}{
        l 
        c 
        c 
        c 
        c 
        c 
        c 
    }
    \toprule
    \textbf{} & \multicolumn{2}{c}{\textbf{Similarity to $q_0$}} & \multicolumn{1}{c}{\textbf{\specificityMetricCap}} & \multicolumn{3}{c}{\textbf{Representativeness}} \\
    \cmidrule(lr){2-3}\cmidrule(lr){4-4}\cmidrule(lr){5-7}
    \textbf{Method} & \makecell{\textbf{CLIP}} & \textbf{Jaccard} & \makecell{\textbf{Recall} \\ \textbf{Cluster}} & \makecell{\textbf{Recall}} & \textbf{NDCG} & \textbf{MAP} \\ 
    
    \midrule
    GroupCap$_\text{Mistral}$ & 0.81 \unc $\pm$ 0.05 & 0.24 \unc $\pm$ 0.09 & \textbf{0.46 \unc $\pm$ 0.19} & 0.43 \unc $\pm$ 0.13 & \textbf{0.35 \unc $\pm$ 0.16} & 0.31 \unc $\pm$ 0.12 \\
    GroupCap$_\text{LLama3}$ & \textbf{0.90 \unc $\pm$ 0.04} & \textbf{0.56 \unc $\pm$ 0.12} & 0.39 \unc $\pm$ 0.17 & \textbf{0.47 \unc $\pm$ 0.12} & \textbf{0.35 \unc $\pm$ 0.18} & \textbf{0.32 \unc $\pm$ 0.14} \\

    \bottomrule
    \end{tabular}
    }
\end{table}

\subsection{Ablation Study}
\subsubsection{Prototype Selection}
We experimented with two ways to select the prototype $\hat{p}$ of the cluster $C$ in CLIP space as input for the prototype captioning methods: a) selecting the \textit{centroid} of the image cluster
\begin{equation}
    \hat{p} = \frac{1}{|C|} \sum_{I \in C} \text{CLIP}(I)\,,
\end{equation}
and b) selecting the most \textit{representative} element as the one having the highest mean cosine similarity to the other images in CLIP space:
\begin{equation}
\label{eq:representativeness}
\hat{p} = \arg\max_{I \in C} \sum_{I' \in C} \cos(\text{CLIP}(I), \text{CLIP}(I'))
\end{equation}
We distinguish the two versions of the same method through the notations $\text{DeCap}_\text{centroid}$ and $\text{DeCap}_\text{repr}$, and $\text{ClipCap}_\text{centroid}$ and $\text{ClipCap}_\text{repr}$ respectively.
\sloppy Considering the scores obtained by the same captioning method (i.e., $\text{DeCap}_\text{centroid}$ versus $\text{DeCap}_\text{repr}$), we can observe that applying the methods over the clusters' centroids generally leads to higher scores. 
In fact, $\text{DeCap}_\text{centroid}$ beats its version applied on representatives in all the metrics, and we reported it in the comparison of Table \ref{table:best-methods-scores-all-metrics} for the \textit{DeCap} method.

\subsubsection{ClipCap: Query Aware vs Vanilla}

Table \ref{table:captioning-scores-all-metrics} shows that the query-aware versions of ClipCap (ClipCap$^{q_0}$) achieve higher scores both in Similarity to Initial Query and Representativeness metrics with respect to vanilla ClipCap, but a lower \specificityMetricCap{}. 
However, query awareness generally leads these models to obtain more balanced scores among the measured properties, which is usually preferred.
For this reason, we picked the query-aware variant as the best configuration we reported in Table \ref{table:best-methods-scores-all-metrics} for ClipCap.

\subsubsection{LLM backbone}

Table \ref{table:groupcap-scores-all-metrics} reports the metrics of GroupCap when using different LLM models for caption summarization.
We tested Mistral-7B~\cite{jiang2023mistral7b} and LLama3-8B~\cite{dubey2024llama} in the GroupCap architecture using DeCap as the captioning model.
\sloppy While GroupCap$_\text{Mistral}$ builds more cluster-specific suggestions, GroupCap$_\text{LLama3}$ achieves the best results both on the Representativeness measures and on Similarity to Initial Query measures, where it even gets a slightly better macro-averaged score than the human-annotated suggestions due to its high capabilities of understanding the few-shot prompt. 
We picked LLama3 as the LLM backbone for the GroupCap configuration reported in Table \ref{table:best-methods-scores-all-metrics}.

\section{Conclusions}
In this paper, we formalized the cross-modal query suggestion task, introducing a set of suitable properties and metrics for quantitatively evaluating different query suggestion methods. We proposed the \benchmarkAcronym{} benchmark, composed of a set of 50 diverse queries and a total of 295 image clusters each having a human-annotated query suggestion.
We adapted two different classes of methods --- two captioning methods and one LLM-based method --- as 
baselines for this novel task. We observed that all the methods achieve important gains over the results obtained using just the initial query, meaning that the proposed baselines can effectively suggest queries 
that better discriminate the clusters,
without diverging too much from the original query formulation. 
However, the gap with human performance suggests that there is still a large margin for improvement in future works.

\begin{credits}
\subsubsection{\ackname}
This work was partially supported by
FAIR -- Future Artificial Intelligence Research - Spoke 1 (PNRR M4C2 Inv. 1.3 PE00000013)
and
MUCES -- a MUltimedia platform for Content Enrichment and Search in audiovisual archives
(PRIN 2022 PNRR P2022BW7CW - CUP: B53D23026090001)
funded by the NextGenerationEU, 
by the Spoke ``FutureHPC \& BigData'' of the ICSC – Centro Nazionale di Ricerca in High-Performance Computing, Big Data and Quantum Computing funded by the Italian Government, the FoReLab and CrossLab projects (Departments of Excellence), the NEREO PRIN project (Research Grant no. 2022AEFHAZ) funded by the Italian Ministry of Education and Research (MUR).
\end{credits}

%
%
%
%
\bibliographystyle{splncs04}
\bibliography{biblio}

\end{document}